\documentclass[%
 reprint,
 amsmath,amssymb,
 aps,
]{revtex4-2}
\usepackage{mathptmx}
\usepackage{etoolbox}
\usepackage{xcolor}
\usepackage{float}
\usepackage{hyperref}
\usepackage{graphicx}
\usepackage{dcolumn}
\usepackage{bm}

\begin{document}

\preprint{APS/123-QED}

\title{Phase Stability and Anisotropic Superconductivity in Ternary M$_2$B$_4$C$_2$ (M=Ca, Mg) Monolayer}

\author{Jakkapat Seeyangnok$^{1}$}
 \email{jakkapatjtp@gmail.com} 
\author{Udomsilp Pinsook$^{2}$}%
\author{Graeme J Ackland$^{3}$}

\affiliation{$^{1}$Department of Physics, Faculty of Science, Chulalongkorn University, Bangkok, Thailand.\\
$^{2}$Centre for Science at Extreme Conditions, School of Physics and Astronomy, University of Edinburgh, Edinburgh, United Kingdom}%

\date{\today}

\begin{abstract}
In this work, we investigate the phase stability, electronic structure, and superconducting properties of the two-dimensional ternary borocarbides $M\text{B}_4\text{C}_2$ ($M = \text{Ca}, \text{Mg}$) using first-principles calculations and fully anisotropic Migdal-Eliashberg theory. Ab initio molecular dynamics, phonon dispersion, and elastic constant calculations confirm that both monolayers are dynamically, thermally, and mechanically stable. Electronic structure analysis reveals an intrinsic metallic nature with a multiband Fermi surface, where states near the $\Gamma$ point are dominated by in-plane B-$p_x, p_y$ orbitals, while zone-boundary states are primarily derived from out-of-plane B-$p_z$ orbitals. The electron-phonon coupling is determined to be $\lambda = 0.64$ for both materials, driven by the rigid covalent B–B and B–C honeycomb networks. By solving the anisotropic Eliashberg equations, we predict that $\text{CaB}_4\text{C}_2$ and $\text{MgB}_4\text{C}_2$ are nodeless two-gap superconductors with critical temperatures ($T_c$) of 33.4 K and 33.8 K, respectively, at a Coulomb pseudopotential of $\mu^* = 0.10$. Furthermore, the superconducting state remains robust under increased Coulomb repulsion ($\mu^* = 0.13$), retaining elevated $T_c$ values of 31.2 K and 30.9 K, respectively. These findings establish $M\text{B}_4\text{C}_2$ monolayers as promising candidates for nanoscale superconducting applications and enrich the family of two-dimensional multigap superconductors.
\end{abstract}

\maketitle

\section{Introduction}
Two-dimensional (2D) materials have attracted tremendous interest owing to their rich physical properties and broad potential in nanoelectronics, energy devices, and quantum technologies~\cite{novoselov2004electric,geim2013van,novoselov20162d,xu2013graphene,manzeli20172d}. Within this landscape, 2D superconductors are of particular interest, because reduced dimensionality can strengthen electronic correlations and electron--phonon interactions, potentially giving rise to unconventional pairing and novel quantum phases~\cite{uchihashi2017two,saito2016superconductivity,xi2016ising}. Phonon-mediated superconductivity in 2D systems is well described by Migdal--Eliashberg theory, in which the interplay between electronic structure and lattice vibrations determines $T_c$~\cite{migdal1958interaction,eliashberg1960interactions,giustino2017electron}. Doped graphene~\cite{profeta2012phonon}, borophene~\cite{gao2017prediction,penev2012polymorphism}, and transition metal dichalcogenides~\cite{ge2015superconductivity,ugeda2016characterization} have been predicted or measured to superconduct with a range of coupling strengths. Systems with complex Fermi surfaces additionally exhibit multiband superconductivity and anisotropic gaps, as exemplified by MgB$_2$~\cite{souma2003origin,liu2001beyond}.

MgB$_2$ is the prototype of this behavior. Its discovery, with $T_c = 39$~K~\cite{nagamatsu2001superconductivity}, established the metal diborides as conventional high-$T_c$ superconductors, and its two-gap state arises from strongly coupled in-plane $\sigma$ states and weakly coupled out-of-plane $\pi$ states. This has motivated extensive efforts to realize 2D analogues. Free-standing few-layer and monolayer MgB$_2$ show a reduced $T_c$ relative to the bulk~\cite{bekaert2017free}, while high-quality thick films reach up to 36~K~\cite{cheng2018fabrication}. Beyond MgB$_2$, first-principles studies have predicted superconductivity in a growing set of 2D metal borides: MB$_4$ (M = Be, Mg, Ca, Al) with $T_c$ up to 36.1~K for CaB$_4$~\cite{sevik2022high} and their hydrogenated derivatives~\cite{seeyangnok2026stability}, MB$_3$ (M = Be, Ca, Sr) and hydrogenated CaB$_3$~\cite{yang2023theoretical}, and tunable MB$_6$ networks~\cite{wan2025coexistence}. Hydrogenation of the MgB$_2$ monolayer has also been shown to raise the predicted $T_c$ substantially~\cite{bekaert2019hydrogen}.

Boron forms electron-deficient networks with strong covalent and multicenter bonding~\cite{oganov2009ionic,pancharatna2022anatomy}, which makes it an effective host for strong electron--phonon coupling. Surface functionalization such as hydrogenation has been widely used to exploit this, including in Janus MoSH, WSH, and Ti$_2$CSH~\cite{lu2017janus,liu2022two,pinsook2025superconductivity,jseeyang_ti2csh,seeyangnok2024superconductivity}. Such systems, however, are often limited by competing charge-density-wave or magnetic instabilities~\cite{ku2023ab,jseeyang_cdw_moxh,seeyangnok2025competition,sukserm2025half}. Incorporating carbon directly into the boron lattice to form ternary borocarbides offers an intrinsic alternative: it tunes the electronic structure and passivates the surfaces without any external functionalization.

A notable realization of this idea is the Mg$_2$B$_4$C$_2$ monolayer proposed by previous study~\cite{singh2022high}. It is obtained by replacing the chemically reactive outer B--B layers of an MgB$_2$ slab with chemically inert B--C layers. This yields a stable, easily exfoliable monolayer with inert surfaces, Dirac-like band crossings near the Fermi level, and a large electron--phonon coupling constant ($\lambda = 1.40$) with estimated $T_c \approx 47$--$48$~K using the isotropic McMillan--Allen-Dynes formula, among the highest predicted intrinsic $T_c$ values for a 2D phonon-mediated superconductor. However, the predicted $T_c$ could vary moderately under a fully anisotropic Migdal--Eliashberg or SC-DFT treatment, and that the applicability of the McMillan--Allen--Dynes formula becomes limited at large electron--phonon coupling~\cite{singh2022high}. The momentum-resolved gap structure, and the role of orbital character in the pairing, therefore remain to be established. The combination of boron and carbon is also expected to produce distinct mixed bonding and multi-orbital character at the Fermi level, which favors multigap superconductivity as seen in n-doped graphene~\cite{margine2014two}, AlB$_2$-based films~\cite{zhao2019two}, trilayer LiB$_2$C$_2$~\cite{gao2020strong}, and monolayer LiBC~\cite{modak2021prediction}.

In this work, we study the structural, electronic, and superconducting properties of the ternary borocarbide monolayers $M\mathrm{B}_4\mathrm{C}_2$ ($M = \mathrm{Ca}, \mathrm{Mg}$) using first-principles calculations and fully anisotropic Migdal--Eliashberg theory. We show that both monolayers are dynamically, thermally, and mechanically stable and intrinsically metallic, with multiband Fermi surfaces in which in-plane B-$p_{x,y}$ states dominate near $\Gamma$ and out-of-plane B-$p_z$ states dominate near the zone boundary. By solving the momentum-resolved Eliashberg equations, we find strongly orbital-dependent electron--phonon coupling that produces a nodeless two-gap superconducting state with $T_c$ above 33~K, which remains above 30~K at a larger Coulomb pseudopotential. These results extend the family of 2D boride superconductors and clarify how orbital hybridization drives anisotropic multigap superconductivity in low-dimensional borocarbides.

\section{Computational Methods}
Density functional theory (DFT) calculations were performed using \textsc{Quantum ESPRESSO}~\cite{giannozzi2009quantum}, employing optimized norm-conserving Vanderbilt pseudopotentials~\cite{hamann2013optimized,schlipf2015optimization} and the Perdew--Burke--Ernzerhof (PBE) generalized gradient approximation~\cite{perdew1996generalized}. Wavefunction and charge density cutoffs were set to 80~Ry and 320~Ry, respectively. The Brillouin zone was sampled with a $24 \times 24 \times 1$ Monkhorst--Pack \textit{k}-mesh~\cite{monkhorst1976special}, utilizing a Methfessel--Paxton smearing of 0.02~Ry~\cite{methfessel1989high}. Structures were relaxed via the BFGS algorithm~\cite{liu1989limited} until residual forces dropped below $10^{-5}$~eV/\AA. Phonon dispersions were computed using density functional perturbation theory (DFPT)~\cite{baroni2001phonons} on a $12 \times 12 \times 1$ \textit{q}-grid.

The mode-resolved electron--phonon coupling strength is defined as:
\begin{equation}
    \lambda_{\boldsymbol{q}\nu} = \frac{\gamma_{\boldsymbol{q}\nu}}{\pi N(\epsilon_F)\,\omega_{\boldsymbol{q}\nu}^2},
\end{equation}
where $\gamma_{\boldsymbol{q}\nu}$ is the phonon linewidth, $N(\epsilon_F)$ is the Fermi-level density of states, and $\omega_{\boldsymbol{q}\nu}$ is the phonon frequency.

To evaluate superconducting properties, we solved the anisotropic Migdal--Eliashberg equations~\cite{migdal1958interaction,eliashberg1960interactions,nambu1960quasi,pinsook2024analytic} via Wannier--Fourier interpolation using the EPW code~\cite{noffsinger2010epw,ponce2016epw,giustino2007electron,giustino2017electron}. The mass renormalization $Z_{nk}(i\omega_j)$ and superconducting gap $\Delta_{nk}(i\omega_j)$ were obtained self-consistently on the Matsubara axis ($\omega_j = (2j+1)\pi T$):
    \begin{eqnarray}
    Z_{nk}(i\omega_j) &=& 1 + \frac{\pi T}{N(\varepsilon_F)\omega_j} 
    \sum_{mk'j'} 
    \frac{\omega_{j'}}{\sqrt{\omega_{j'}^2 + \Delta_{mk'}^2(i\omega_{j'})}} \nonumber \\
    &&\times \lambda(nk, mk', \omega_j - \omega_{j'})\,
    \delta(\epsilon_{mk'} - \varepsilon_F),
    \end{eqnarray}
    \begin{eqnarray}
    Z_{nk}(i\omega_j)\Delta_{nk}(i\omega_j) &=& 
    \frac{\pi T}{N(\varepsilon_F)} 
    \sum_{mk'j'} 
    \frac{\Delta_{mk'}(i\omega_{j'})}{\sqrt{\omega_{j'}^2 + \Delta_{mk'}^2(i\omega_{j'})}}  \delta(\epsilon_{mk'} - \varepsilon_F) \nonumber  \\ &&
    \times \left[\lambda(nk, mk', \omega_j - \omega_{j'}) - \mu^*\right],
    \end{eqnarray}
We tested Coulomb pseudopotentials of $\mu^* = 0.1$ and $0.13$. Convergence was achieved using dense interpolation grids ($160 \times 160 \times 1$ for \textit{k}-points and $80 \times 80 \times 1$ for \textit{q}-points), a 1.00~eV Matsubara cutoff, and Gaussian smearing widths of 0.40~eV for electrons and 0.5~meV for phonons.

Finally, Ab Initio Molecular Dynamics (AIMD) simulations were conducted on a $2 \times 2 \times 1$ supercell in the NVT ensemble (300~K) with a 1.0~fs time step. A Nosé--Hoover thermostat~\cite{NHC} was applied with a 1.0~ps coupling time constant, which corresponds directly to the $\sim 1.0$~ps macroscopic energy oscillation period observed during the simulation.

\section{Results and Discussion}

\subsection{Crystal Structures and Phase Stability}
\begin{figure}[htbp]
    \centering
    \includegraphics[width=8.5cm]{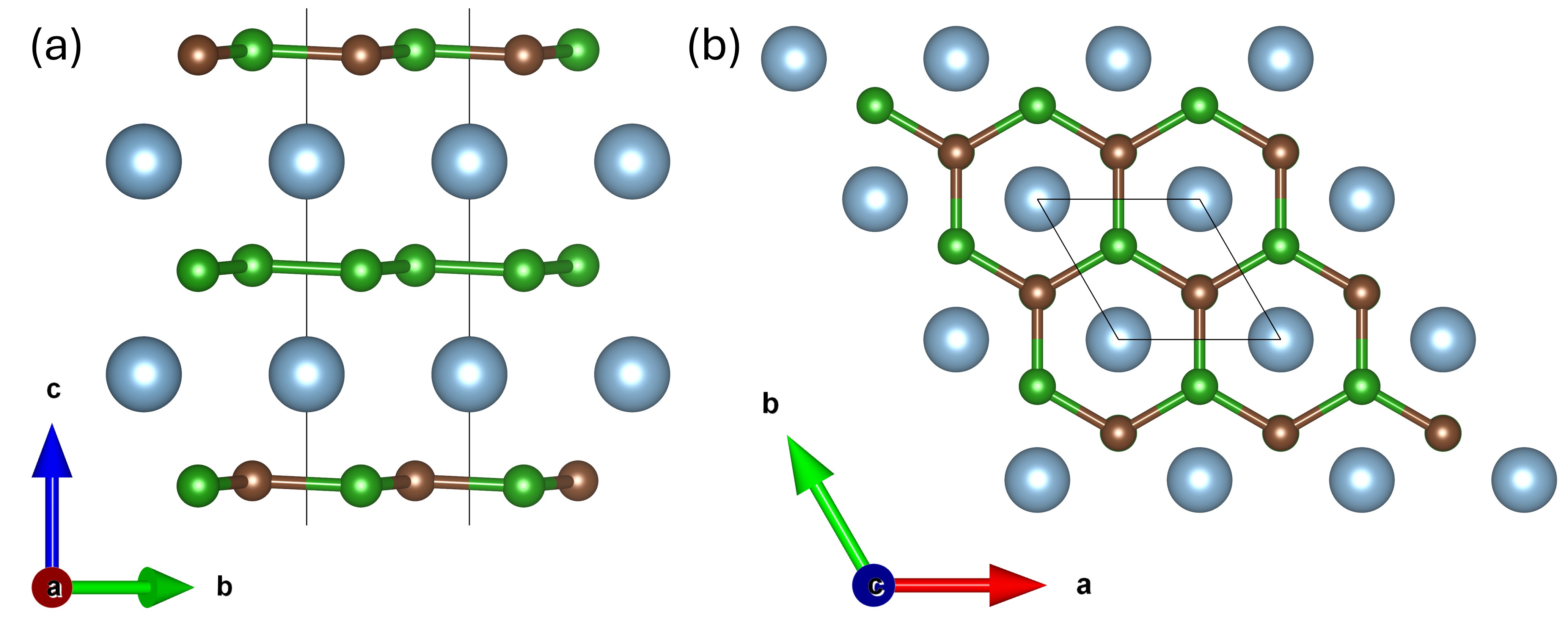}
    \caption{(a) Side view and (b) top view of the crystal structure of $M\text{B}_4\text{C}_2$ ($M$ =Ca, Mg) in space group $P3m1$ (No. 156). Light blue, green, and brown spheres represent $M$, B, and C atoms, respectively.}
    \label{fig:structure}
\end{figure}

The $M\text{B}_4\text{C}_2$ monolayers ($M = \text{Ca}, \text{Mg}$) crystallize in the $P3m1$ space group (No. 156), as shown in Fig. \ref{fig:structure}. The atomic structure consists of alternating boron and carbon sublayers with intercalated metal $M$ layers. Within the unit cell, the metal atoms occupy the $1a$ Wyckoff positions at $(0, 0, z_M)$, while the carbon and boron atoms are located at the $1b$ and $1c$ positions at $(1/3, 2/3, z)$ and $(2/3, 1/3, z)$, respectively, forming the stacked hexagonal network. 

\begin{table}[htbp]
  \centering
  \caption{Calculated in-plane lattice constants $a$ (\AA), interlayer spacing between carbon--boron and boron--boron sublayers ($d_{\text{CB--BB}}$ in \AA), and vertical clearances from the metal layer to the adjacent boron--boron ($d_{M\text{--BB}}$ in \AA) and carbon--boron sublayers ($d_{M\text{--CB}}$ in \AA) for monolayer $M\text{B}_4\text{C}_2$ ($M = \text{Ca}, \text{Mg}$) in space group $P3m1$ (No. 156).}
  \label{tab:latticeparameters}
  \begin{tabular}{lcccc}
    \hline\hline
    System & $a$ & CB--BB Layer & $M$--BB Layer & $M$--CB Layer \\
    \hline
    CaB$_4$C$_2$ & 2.95 & 4.05 & 2.07 & 2.12 \\
    MgB$_4$C$_2$ & 2.68 & 3.60 & 1.79 & 1.81 \\
    \hline\hline
  \end{tabular}
\end{table}

The lattice parameter $a$ scales directly with the atomic radius of the intercalated metal. Among the three systems, $\text{MgB}_4\text{C}_2$ exhibits the smallest lattice constant at $a = 2.68$~\AA, whereas $\text{CaB}_4\text{C}_2$ has the largest at $a = 2.95$~\AA. The vertical layer spacing follows the same trend. The distance between the CB and BB sublayers expands from $3.60$~\AA\ in $\text{MgB}_4\text{C}_2$ to $4.05$~\AA\ in $\text{CaB}_4\text{C}_2$. Across all three systems, the metal layer sits slightly closer to the BB sublayer than to the CB sublayer, as summarized in Table~\ref{tab:latticeparameters}.

    \begin{table}[htbp]
  \centering
  \caption{Calculated 2D elastic constants ($C_{ij}$, in N/m) for the hexagonal $M\text{B}_4\text{C}_2$ ($M = \text{Al}, \text{Ca}, \text{Mg}$) monolayers.}\label{tab:elasticconstant}
  \label{tab:elastic}
  \begin{tabular}{lcccc}
    \hline\hline
    System & $C_{11}=C_{22}$ & $C_{12}$ & $C_{66}$ \\
    \hline
    $\text{CaB}_4\text{C}_2$ & 440.76 & 109.25 & 165.75 \\
    $\text{MgB}_4\text{C}_2$ & 613.81 & 87.06 & 263.38 \\
    \hline\hline
  \end{tabular}
\end{table}


We evaluate mechanical stability via in-plane elastic constants ($C_{ij}$), derived from the strain--energy relationships of the relaxed structure:

\begin{equation}
    C_{ij} = \frac{1}{S_0} \frac{\partial^2 E}{\partial \epsilon_i \partial \epsilon_j}
\end{equation}
where $S_0$ is the equilibrium unit cell area, and $\epsilon_i, \epsilon_j$ are the applied planar strain components. For hexagonal symmetry, the elastic tensor reduces to two independent constants, $C_{11}$ and $C_{12}$, with the shear modulus $C_{66} = (C_{11} - C_{12})/2$. The 2D generalized Hooke's law is:
\begin{equation}
    \sigma = \begin{bmatrix} C_{11} & C_{12} & 0 \\ C_{12} & C_{11} & 0 \\ 0 & 0 & C_{66} \end{bmatrix} \varepsilon.
\end{equation}

The calculated elastic constants for the $M\text{B}_4\text{C}_2$ ($M =\text{Ca}, \text{Mg}$) monolayers are summarized in Table~\ref{tab:elasticconstant}. $\text{MgB}_4\text{C}_2$ exhibits the highest stiffness ($C_{11} = 613.81$~N/m, $C_{12} = 87.06$~N/m, $C_{66} = 263.38$~N/m), followed by $\text{CaB}_4\text{C}_2$ ($C_{11} = 440.76$~N/m, $C_{12} = 109.25$~N/m, $C_{66} = 165.75$~N/m). All three systems strictly satisfy the Born criteria for 2D mechanical stability ($C_{11}C_{22} - C_{12}^2 > 0$ and $C_{11}, C_{22}, C_{66} > 0$)~\cite{mouhat2014necessary}, confirming their mechanical robustness as free-standing monolayers.

\begin{figure}[htbp]
    \centering
    \includegraphics[width=8.5cm]{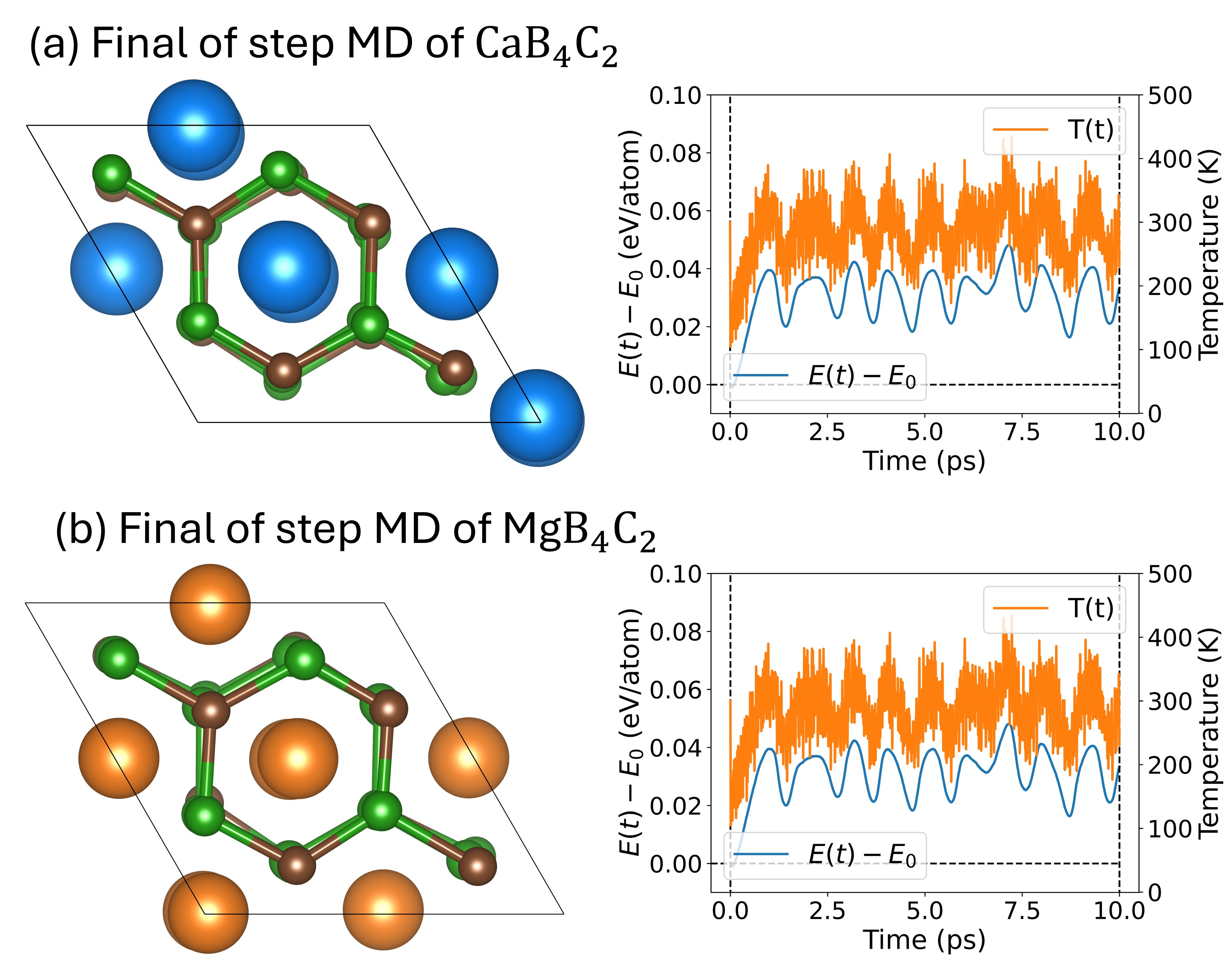}
    \caption{Ab initio molecular dynamics (AIMD) simulations of the $M\text{B}_4\text{C}_2$ monolayers at 300 K for 10 ps. The results are shown for (a) $\text{CaB}_4\text{C}_2$, and (b) $\text{MgB}_4\text{C}_2$. In each panel, the left side displays the final structural snapshot of the $2 \times 2$ supercell at the end of the simulation. The right side shows the time evolution of the temperature $T(t)$ (orange line, right axis) and the total energy variation per atom $E(t) - E_0$ (blue line, left axis).}
    \label{fig:aimd}
\end{figure}    

To evaluate the thermal stability of the $M\text{B}_4\text{C}_2$ monolayers at room temperature, we performed AIMD simulations at 300 K for a duration of 10 ps. Figure~\ref{fig:aimd} presents the final atomic configurations at the end of the simulation alongside the thermodynamic profiles. The structural snapshots reveal that the original hexagonal lattice framework remains intact for all three systems ($\text{CaB}_4\text{C}_2$, and $\text{MgB}_4\text{C}_2$). The atoms exhibit normal thermal displacements around their equilibrium sites, with no evidence of bond breakage, severe geometric distortion, or phase reconstruction. 

The thermodynamic data further support this thermal stability. The right panels of Fig.~\ref{fig:aimd} track the temperature and the total energy per atom relative to the initial state ($E(t) - E_0$). Throughout the 10 ps timeframe, the temperature fluctuates consistently around the target value of 300 K. The total energy remains steady, with variations confined to a small window (less than $\sim 0.06$ eV/atom) and no indication of a continuous upward or downward drift. As noted in the methodology, the $\sim 1.0$ ps periodic oscillations observed in the energy profiles correspond directly to the coupling time constant of the Nosé-Hoover thermostat. The combination of structural preservation and sustained thermodynamic equilibrium confirms that the $M\text{B}_4\text{C}_2$ monolayers are thermally stable at room temperature.
\subsection{Electronic Properties}
    \begin{figure}[ht]
        \centering
        \includegraphics[width=8.5cm]{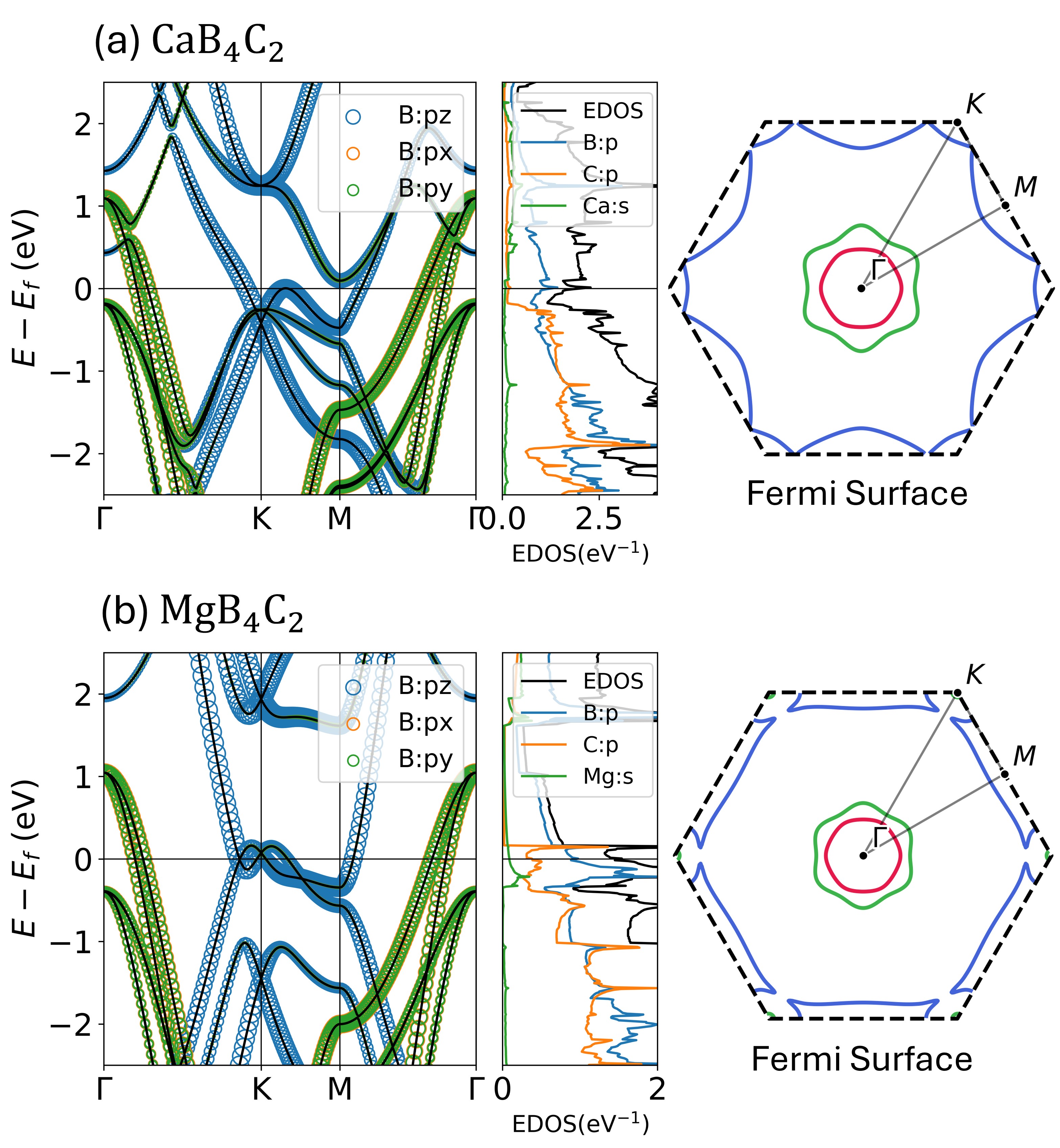}
        \caption{Electronic band structures, projected density of states (PDOS), and Fermi surfaces for monolayer (a) $\text{AlB}_4\text{C}_2$, (b) $\text{CaB}_4\text{C}_2$, and (c) $\text{MgB}_4\text{C}_2$. In the band structure plots, orbital contributions are highlighted by open circles, where the marker size corresponds to the spectral weight of the specified state. The horizontal solid line at zero energy indicates the Fermi level ($E_F$). The dashed black hexagons represent the first Brillouin zone for the Fermi surfaces.}
    \label{fig:electronic}
    \end{figure}
    
    The electronic band structures, projected density of states (PDOS), and Fermi surfaces for the $M\text{B}_4\text{C}_2$ monolayers ($M = \text{Ca}, \text{Mg}$) are presented in Fig.~\ref{fig:electronic}. Both materials exhibit intrinsic metallic behavior, characterized by highly dispersive bands crossing the Fermi level ($E_F$).

    A detailed analysis of the orbital projections reveals a distinct momentum-dependent orbital character for the electronic states in the vicinity of $E_F$. As highlighted by the projected band structure plots, the bands intersecting the Fermi level near the Brillouin zone center ($\Gamma$ point) are primarily composed of the in-plane boron $p_x$ and $p_y$ states. Conversely, the highly dispersive bands crossing $E_F$ near the zone boundaries (along the $K$ and $M$ paths) originate almost entirely from the out-of-plane boron $p_z$ orbitals (blue circles). This orbital decomposition is corroborated by the PDOS, which shows that the total density of states at $E_F$ is chiefly derived from the overall $\text{B-}p$ states. Contributions from the carbon $p$ orbitals and the intercalated metal states ($\text{Ca-}s$, $\text{Mg-}s$) are negligible at the Fermi level, with their spectral weights located deeper in the valence or conduction regions.

    The topology of the electronic structure is highly dependent on the choice of the intercalated metal, particularly regarding a prominent B-$p_z$ band intersection located near the $K$ high-symmetry point. The energetic position of this crossing relative to $E_F$ directly dictates the shape of the outer Fermi surface. In $\text{MgB}_4\text{C}_2$, this specific band intersection aligns directly with the Fermi energy. As a result, the Fermi surface reflects this crossing through the formation of small, disconnected pockets along the outer hexagonal Brillouin zone boundary centered near the $K$ points. In contrast, for $\text{CaB}_4\text{C}_2$, the intersection is slightly offset from $E_F$, merging these disconnected pockets into a continuous, undulating outer contour.

    \begin{figure}[ht]
        \centering
        \includegraphics[width=8.5cm]{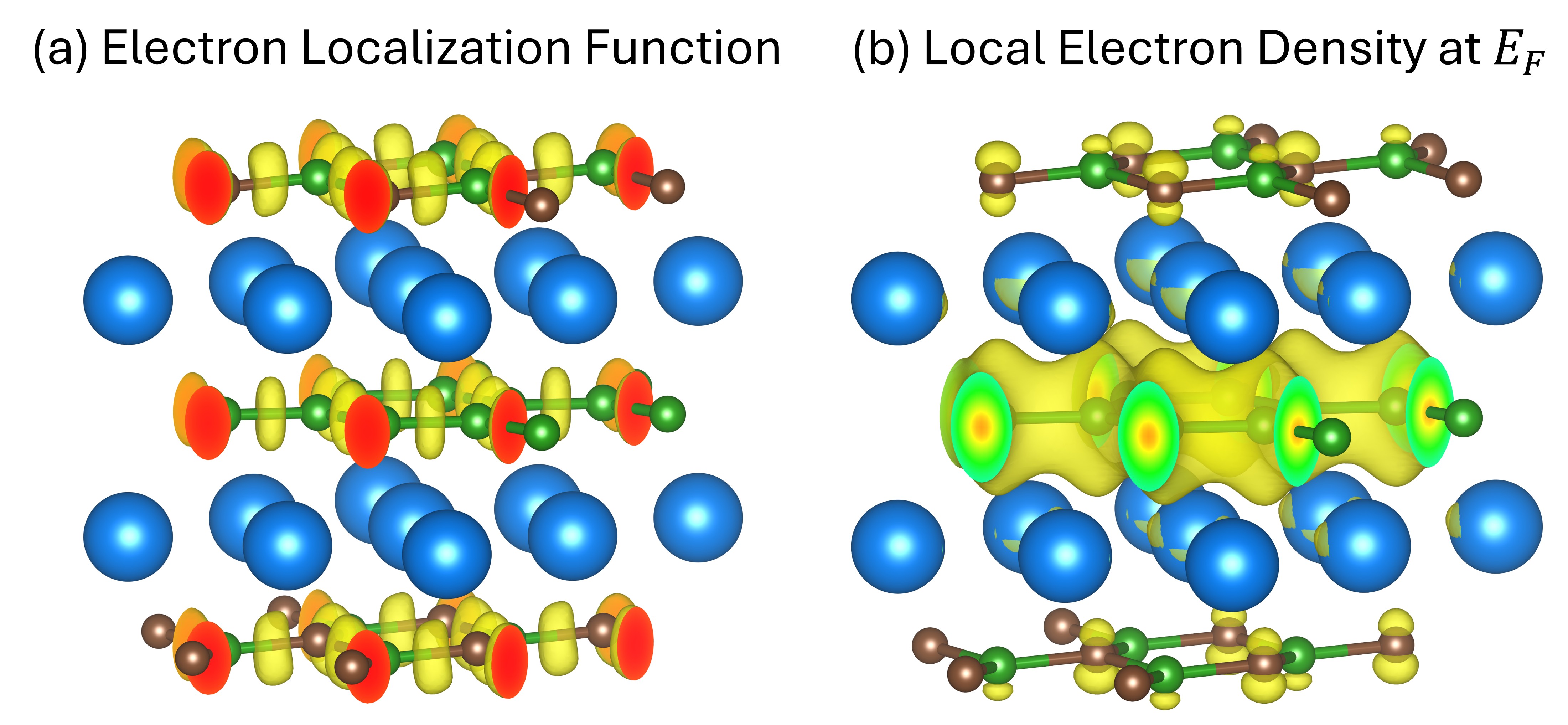}
        \caption{Three-dimensional spatial distributions of (a) the electron localization function (ELF) and (b) the local electron density evaluated at the Fermi level ($E_F$) for the $M$B$_4$C$_2$ systems. The ELF highlights the strong covalent bonds within the honeycomb network, while the local density at $E_F$ visualizes the $p$-orbital character of the conduction electrons.}
    \label{fig:electron_density}
    \end{figure}

    To further understand the chemical bonding and its relation to the lattice dynamics, the electron localization function (ELF) is visualized in Figure~\ref{fig:electron_density}(a). The ELF reveals highly localized electron distributions situated precisely between the B--B and B--C atoms, acting as strong attractors that indicate robust covalent bonding within the layered honeycomb networks. Consequently, the combination of this strong, rigid covalent bonding and the inherently light atomic masses of boron and carbon explains why the B- and C-projected vibrations overwhelmingly dominate the middle- and high-frequency regions of the phonon spectrum.

    Furthermore, the local electron density evaluated at the Fermi level ($E_F$), shown in Figure~\ref{fig:electron_density}(b), clearly displays the spatial distribution of the active electronic states responsible for the metallic behavior. The charge density at $E_F$ is heavily dominated by distinct $p$-orbital characteristics (visible as the prominent lobes centered on the atoms in the honeycomb network). This spatial visualization perfectly corroborates the projected electronic density of states (PDOS) discussed previously, confirming that the electronic states traversing the Fermi level—and ultimately driving the superconductivity—originate almost exclusively from the boron $p$ orbitals.
\subsection{Phonon and Electron-Phonon Coupling}
    \begin{figure}[ht]
        \centering
        \includegraphics[width=8.5cm]{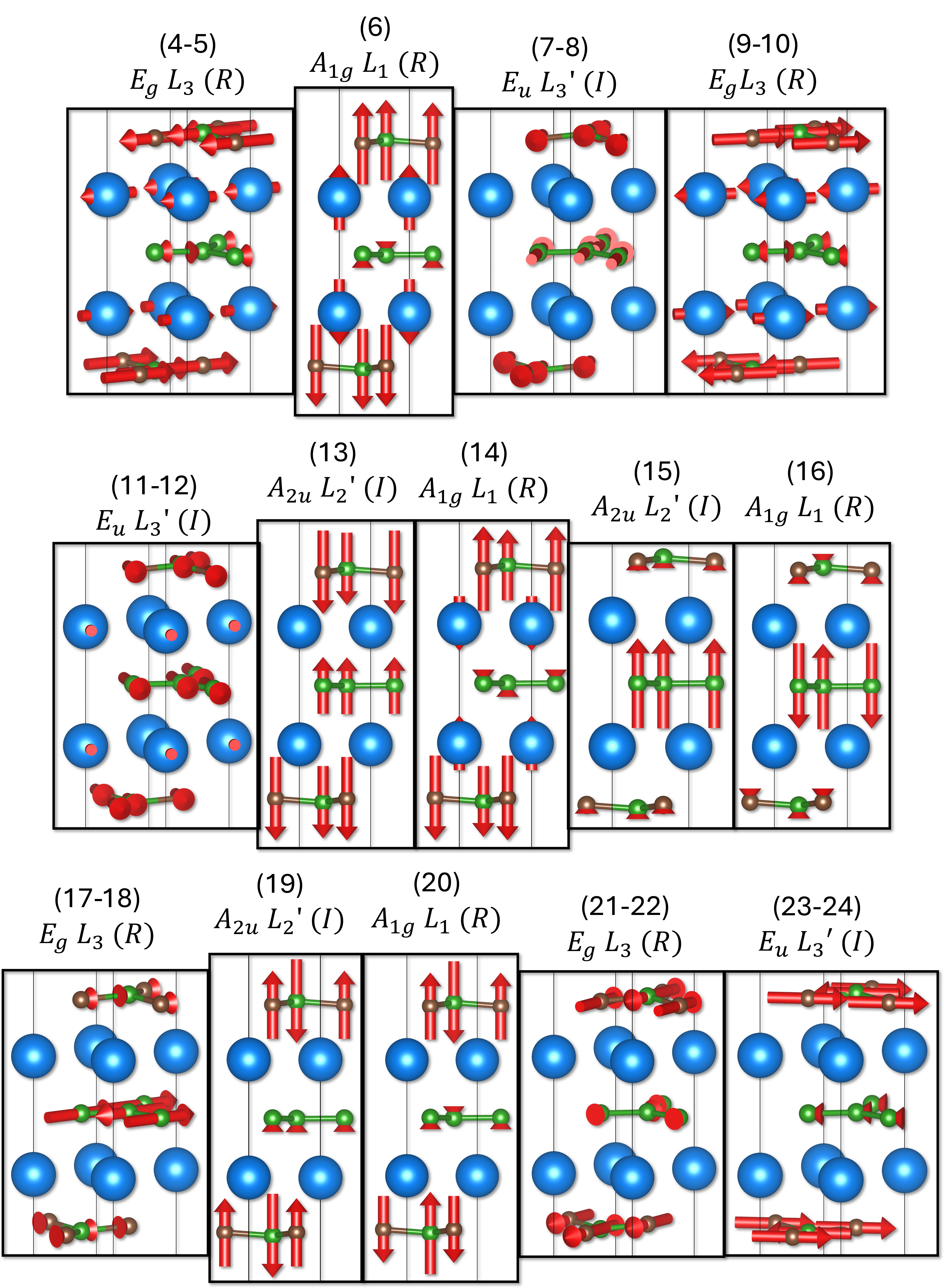}
        \caption{Schematic representation of the atomic displacement patterns for the optical phonon modes at the Brillouin zone center ($\Gamma$ point) of the $M$B$_4$C$_2$ monolayers. Each panel is labeled with the corresponding mode index (based specifically on the CaB$_4$C$_2$ system), symmetry, irreducible representation, and optical activity (R for Raman, I for Infrared). Red arrows indicate the direction and relative amplitude of the atomic vibrations.}
    \label{fig:phonon_modes}
    \end{figure}
    
    \begin{figure}[ht]
        \centering
        \includegraphics[width=8.0cm]{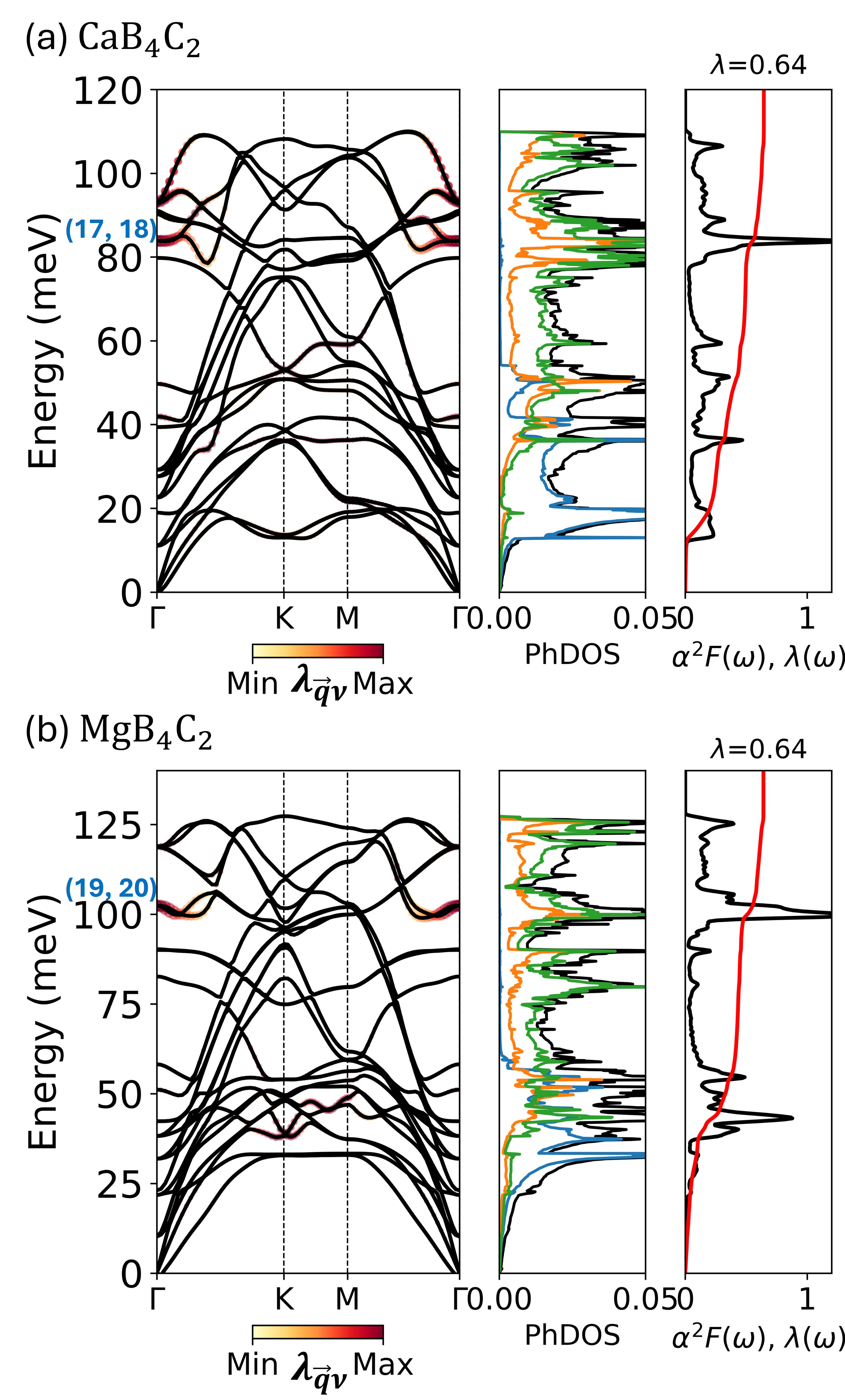}
        \caption{Phonon dispersion with electron-phonon coupling weights (left), atomic-projected phonon density of states (middle), and the Eliashberg spectral function $\alpha^2F(\omega)$ alongside the cumulative coupling constant $\lambda(\omega)$ (right).}
    \label{fig:elph}
    \end{figure}

\begin{table}[htbp]
    \centering
    \caption{Calculated vibrational modes, irreducible representations (Irreps), optical activities (Raman vs. Infrared), and vibrational frequencies (Freq., in meV) at the Brillouin zone center ($\Gamma$ point) for CaB$_4$C$_2$ and MgB$_4$C$_2$, belonging to the $D_{3d}$ ($-3m$) point group symmetry.}
    \label{tab:phonon_gamma}
    \begin{tabular}{cccccc}
        \hline
        \hline
        Symmetry & Activity & Ca Modes & Ca Freq. & Mg Modes & Mg Freq. \\
        \hline
        $E_{g}$ $L_3$  & Raman & 4--5   & 11.38 & 4--5   & 10.66 \\
        $A_{1g}$ $L_1$ & Raman & 6      & 19.16 & 8      & 23.24 \\
        $E_{u}$ $L_3'$ & IR    & 7--8   & 22.61 & 6--7   & 21.90 \\
        $E_{g}$ $L_3$  & Raman & 9--10  & 27.75 & 9--10  & 31.91 \\
        $E_{u}$ $L_3'$ & IR    & 11--12 & 29.25 & 11--12 & 38.18 \\
        $A_{2u}$ $L_2'$ & IR    & 13     & 39.49 & 13     & 42.41 \\
        $A_{1g}$ $L_1$ & Raman & 14     & 42.02 & 14     & 51.17 \\
        $A_{2u}$ $L_2'$ & IR    & 15     & 49.74 & 15     & 58.21 \\
        $A_{1g}$ $L_1$ & Raman & 16     & 79.77 & 16     & 82.65 \\
        $E_{g}$ $L_3$  & Raman & 17--18 & 83.82 & 19--20 & 102.37 \\
        $A_{2u}$ $L_2'$ & IR    & 19     & 90.38 & 17     & 90.05 \\
        $A_{1g}$ $L_1$ & Raman & 20     & 91.16 & 18     & 90.27 \\
        $E_{g}$ $L_3$  & Raman & 21--22 & 92.85 & 21--22 & 118.71 \\
        $E_{u}$ $L_3'$ & IR    & 23--24 & 93.10 & 23--24 & 118.93 \\
        \hline
        \hline
    \end{tabular}
\end{table}

    TThe eigenvalues at the $\Gamma$ point for CaB$_4$C$_2$ and MgB$4$C$2$ are summarized in Table~\ref{tab:phonon_gamma}, and the corresponding atomic displacement patterns for these optical modes are visualized in Figure~\ref{fig:phonon_modes}. As depicted, the vibrational eigenvectors clearly separate into in-plane motions—associated with the doubly degenerate $E_g$ and $E_u$ symmetries—and out-of-plane motions—associated with the non-degenerate $A{1g}$ and $A{2u}$ symmetries. The phonon dispersions along high-symmetry paths are shown in Figure~\ref{fig:elph}. Because the primitive cell contains seven atoms, these structures give rise to 3 acoustic and 21 optical phonon branches (a total of 24 vibrational branches). The absence of imaginary frequencies across the entire Brillouin zone indicates that these materials are lattice-dynamically stable, which, alongside previous studies of their mechanical and thermal properties, indicates robust structural stability. As observed in the dispersion curves, the phonon energies extend up to approximately 110 meV for CaB$_4$C$_2$ and nearly 125 meV for MgB$_4$C$_2$. At the $\Gamma$ point, CaB$_4$C$_2$ notably exhibits a distinct phonon band gap between roughly 50 and 80 meV, cleanly separating the mid- and high-frequency optical branches. In contrast, the phonon branches in MgB$_4$C$_2$ are more continuous throughout this intermediate energy range.

    The atomic-projected phonon density of states (PhDOS) reveals a distinct separation of vibrational characteristics driven by atomic mass. Due to their heavier masses, the alkaline-earth metals dominate the low-frequency spectrum, comprising almost exclusively the acoustic and lowest-lying optical branches (below 55 meV for Ca and 60 meV for Mg). Conversely, the lighter B--B atomic networks hybridize strongly to form rigid two-dimensional structures; their vibrations occupy the middle-frequency range, whereas the modes associated with the B--C atomic networks occupy the high-frequency regions. Together, they dominate the broad PhDOS spectrum from roughly 40 meV up to the maximum phonon energies.

    The role of phonons in the phonon-mediated electron-electron interaction is analyzed through the EPC-weighted phonon dispersion and the Eliashberg spectral function, $\alpha^2F(\omega)$. For the Ca-based material, there is a continuous contribution of glue phonons across the phonon spectrum, leading to a steadily increasing cumulative electron-phonon coupling ($\lambda(\omega)$). A significant contribution—as shown by the peak in the Eliashberg spectral function—comes from the (17, 18) optical modes, which display the in-plane vibrations of the B--B atomic networks. For the Mg-based material, the significant contributions to the electron-phonon coupling ($\lambda$) result from the same in-plane vibrations of the B--B atomic networks associated with the (19, 20) optical modes, as well as the modes around 40--50~meV at the zone boundaries $K$ and $M$, as shown in the EPC-weighted phonon dispersion and the Eliashberg spectral function. The total electron-phonon coupling ($\lambda$) is 0.64 for both the Ca- and Mg-based materials.
    
\subsection{Anisotropic Superconductivity}
    \begin{figure*}[ht]
        \centering
        \includegraphics[width=17.5cm]{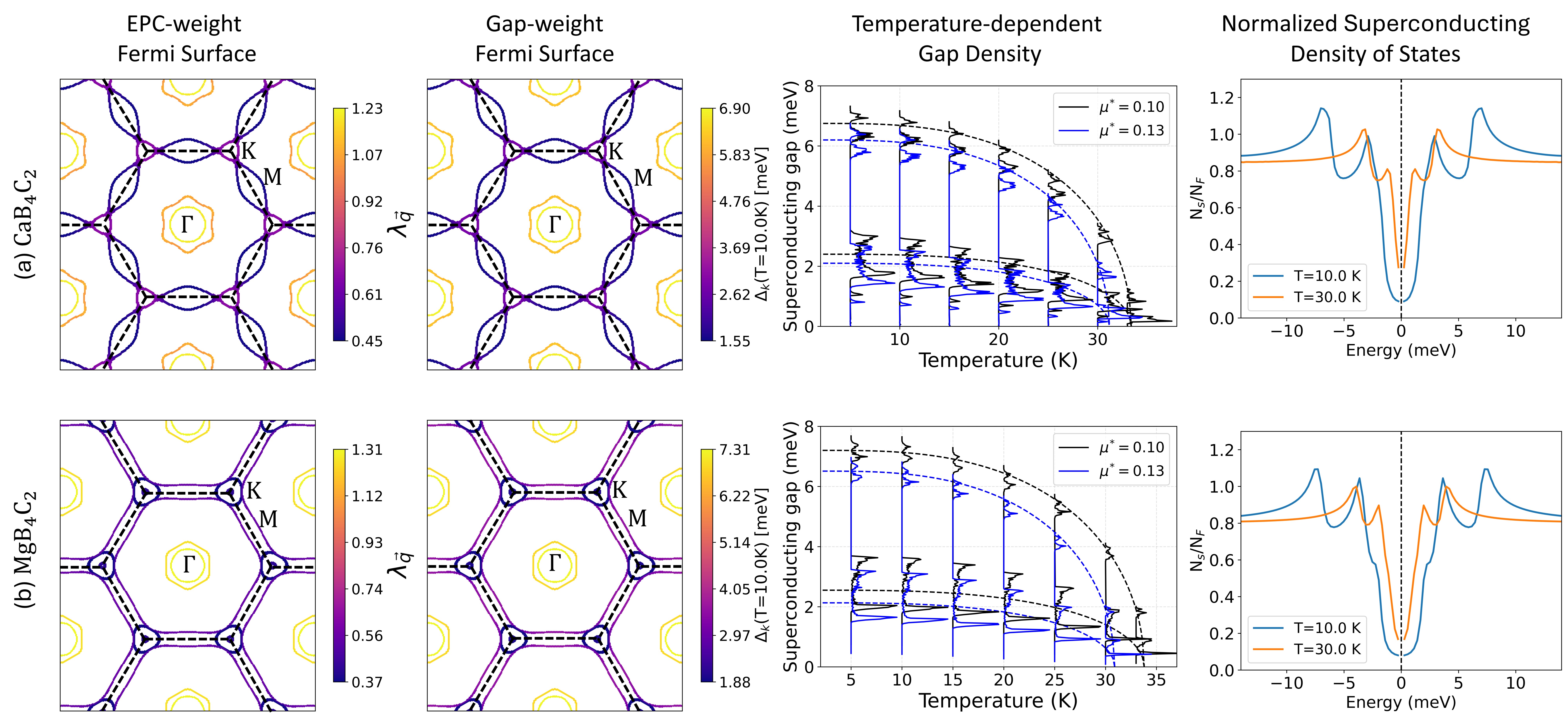}
        \caption{Superconducting properties of CaB$_4$C$_2$ (a) and MgB$_4$C$_2$ (b). From left to right: the momentum-resolved electron-phonon coupling (EPC) parameter ($\lambda_{\mathbf{q}}$) projected onto the Fermi surface; the superconducting energy gap ($\Delta_{\mathbf{k}}$) at $T=10.0$~K mapped onto the Fermi surface; the temperature-dependent anisotropic superconducting gap distributions with BCS fits (dashed lines); and the normalized superconducting quasiparticle density of states ($N_s/N_F$) at $T=10.0$~K and $30.0$~K. All calculations were performed using a Coulomb pseudopotential of $\mu^* = 0.10$, except for the temperature-dependent gap density plots (third column), which include a comparison with $\mu^* = 0.13$ (blue lines).}
    \label{fig:superconductivity}
    \end{figure*}

    The superconducting properties of CaB$_4$C$_2$ and MgB$4$C$2$ were investigated by solving the fully anisotropic Migdal-Eliashberg equations. The momentum-resolved electron-phonon coupling (EPC) parameter ($\lambda{\mathbf{q}}$) and the superconducting energy gap ($\Delta{\mathbf{k}}$) are projected onto their respective Fermi surfaces in the first two columns of Figure~\ref{fig:superconductivity}. Both materials exhibit a highly anisotropic, multiband Fermi surface topology consisting of disconnected pocket-like structures centered around the $\Gamma$ point and an interconnected network spanning towards the K and M points in the hexagonal Brillouin zone. The EPC strength is not uniformly distributed; rather, the inner pockets near the $\Gamma$ point display significantly stronger coupling (yellow regions) compared to the outer networks (purple regions). This behavior is strongly orbital-dependent: the strong EPC and the resulting larger gap around the $\Gamma$ point originate primarily from the in-plane B-$p_x$ and $p_y$ orbitals, while the weaker EPC and smaller gap associated with the edge states are predominantly driven by the out-of-plane B-$p_z$ orbitals. This momentum-dependent coupling directly translates into the gap-weight Fermi surface, where the strongly coupled sheets open a notably larger superconducting gap than the weakly coupled sheets, providing a direct visual signature of multigap superconductivity.

    The temperature evolution of the anisotropic superconducting gap, shown in the third column of Figure~\ref{fig:superconductivity}, explicitly confirms the two-gap nature of both materials. At low temperatures ($T = 5.0$~K), the gap distribution strictly bifurcates into two distinct energy scales. For CaB$_4$C$_2$ at a Coulomb pseudopotential of $\mu^* = 0.10$, the larger gap ($\Delta_2$) settles at 7.2~meV, while the smaller gap ($\Delta_1$) centers at 3.2~meV. MgB$_4$C$_2$ exhibits slightly enhanced values at this temperature, with the larger gap at 7.3~meV and the smaller gap at 3.6~meV. As the temperature increases, both the large and small gaps monotonically decrease and simultaneously close at a single superconducting critical temperature ($T_c$), which is 33.4~K for CaB$_4$C$_2$ and 33.8~K for MgB$_4$C$_2$. This concurrent closing perfectly follows the standard Bardeen-Cooper-Schrieffer (BCS) temperature dependence (black dashed lines), a hallmark of two-gap systems with strong interband scattering. To test the robustness of the superconductivity against electron-electron repulsion, calculations were also performed at $\mu^* = 0.13$. As expected, the heightened Coulomb repulsion suppresses the pairing interaction, reducing $T_c$ to 31.2~K and 30.9~K for the Ca- and Mg-based compounds, respectively. Correspondingly, the low-temperature ($T = 5.0$~K) gap distributions shift to lower energies (blue lines), shrinking to $\Delta_1 = 2.7$~meV and $\Delta_2 = 6.6$~meV for CaB$_4$C$_2$, and $\Delta_1 = 3.1$~meV and $\Delta_2 = 6.5$~meV for MgB$_4$C$_2$.

    It is instructive to compare our results with the earlier prediction for the Mg$_2$B$_4$C$_2$ monolayer ~\cite{singh2022high}, where $\lambda = 1.40$ and $T_c = 47.2$~K at $\mu^{*} = 0.10$ using the isotropic McMillan--Allen--Dynes formula. We obtain a weaker coupling, $\lambda = 0.64$, and a lower $T_c$ of 33.4~K at the same $\mu^{*}$ of 0.10. This difference may arise partly from the methods. As noted by previous study~\cite{singh2022high} that the McMillan--Allen--Dynes formula has limited applicability at large electron--phonon coupling, and the predicted $T_c$ could vary under a fully anisotropic Migdal--Eliashberg treatment. Our calculation resolves the momentum dependence of both the coupling and the gap, whereas the isotropic formula treats pairing as a single channel.

    The normalized superconducting quasiparticle density of states ($N_s/N_F$), presented in the rightmost column, provides further complementary evidence of the multiband superconducting state. At $T = 10.0$~K, the spectra for both materials display pronounced double coherence peaks symmetrically situated around the Fermi energy (0 meV). The inner peaks correspond to the smaller gap on the outer Fermi surface network, while the broader, more prominent outer peaks correspond to the larger gap on the $\Gamma$-centered pockets. The region between the inner peaks exhibits a true zero density of states, confirming fully developed, nodeless $s$-wave gaps. As the temperature is raised to 30.0~K—approaching the critical temperature—thermal broadening smears the quasiparticle excitations. The coherence peaks widen, shift inward, and the gap region begins to fill with states, directly visualizing the thermal destruction of the superconducting condensate just below $T_c$.

    Therefore, these results firmly place the hexagonal $M$B$_4$C$_2$ ($M =$ Ca, Mg) monolayers among a distinguished class of known two-gap superconductors. Their intrinsic multigap nature is highly analogous to the behavior observed in other prominent layered systems, such as n-doped graphene~\cite{margine2014two}, AlB$_2$-based thin films~\cite{zhao2019two}, trilayer LiB$_2$C$_2$~\cite{gao2020strong}, monolayer LiBC~\cite{modak2021prediction}, GaInSLi~\cite{seeyang2025phase_japtwpgap}, MoSLi~\cite{xie2024strong}, MoSeLi~\cite{seeyangnok2026tunable}, MoSH~\cite{liu2022two}, and hydrogenated borides (MgB$_4$H, CaB$_4$H, and AlB$_4$H)~\cite{seeyangnok2026stability}.

\section*{Conclusions}
In summary, we have systematically investigated the structural, electronic, vibrational, and superconducting properties of the hexagonal $M\text{B}_4\text{C}_2$ ($M = \text{Ca}, \text{Mg}$) monolayers through first-principles density functional theory and anisotropic Migdal-Eliashberg calculations. We established that both materials are fully stable in their free-standing 2D forms, satisfying the Born criteria for mechanical stability, lacking imaginary phonon frequencies across the Brillouin zone, and maintaining their structural integrity at 300 K during ab initio molecular dynamics simulations.

Both $\text{CaB}_4\text{C}_2$ and $\text{MgB}_4\text{C}_2$ exhibit intrinsic metallic behavior with highly anisotropic, multiband Fermi surfaces. The electron-phonon interaction is strongly orbital-dependent, yielding a total coupling strength of $\lambda = 0.64$ in both systems. The high-frequency optical phonons, originating from the rigid B–B and B–C covalent honeycomb networks, couple strongly with the in-plane B-$p_x, p_y$ orbitals near the $\Gamma$ point, while the out-of-plane B-$p_z$ orbitals at the zone boundaries exhibit significantly weaker coupling.

This pronounced momentum- and orbital-dependent electron-phonon coupling gives rise to robust two-gap $s$-wave superconductivity. At $T=5.0$ K and a Coulomb pseudopotential of $\mu^* = 0.10$, the superconducting states are characterized by a larger gap ($\Delta_2 \approx 7.2-7.3$ meV) on the $\Gamma$-centered pockets and a smaller gap ($\Delta_1 \approx 3.2-3.6$ meV) on the outer Fermi surface networks. Both energy gaps close concurrently at superconducting critical temperatures ($T_c$) of 33.4 K for $\text{CaB}_4\text{C}_2$ and 33.8 K for $\text{MgB}_4\text{C}_2$. Even under heightened electron-electron repulsion ($\mu^* = 0.13$), the superconductivity remains highly stable, with $T_c$ only moderately suppressed to 31.2 K and 30.9 K for the Ca- and Mg-based monolayers, respectively. Ultimately, these results firmly place the $M\text{B}_4\text{C}_2$ monolayers among an elite class of intrinsic 2D two-gap superconductors, highlighting the critical role of multi-orbital hybridization in driving high-$T_c$ multigap superconductivity in layered borocarbide systems.

    
\section*{Appendix A: Wannier Interpolation Validation}

To accurately evaluate the electron--phonon coupling and solve the Migdal--Eliashberg equations, we utilized Maximally Localized Wannier Functions (MLWFs) to interpolate the electronic states onto ultra-dense momentum grids. The accuracy of this approach relies on how well the Wannier-interpolated bands reproduce the original first-principles data. 

\begin{figure}[htbp]
    \centering
    \includegraphics[width=8.7cm]{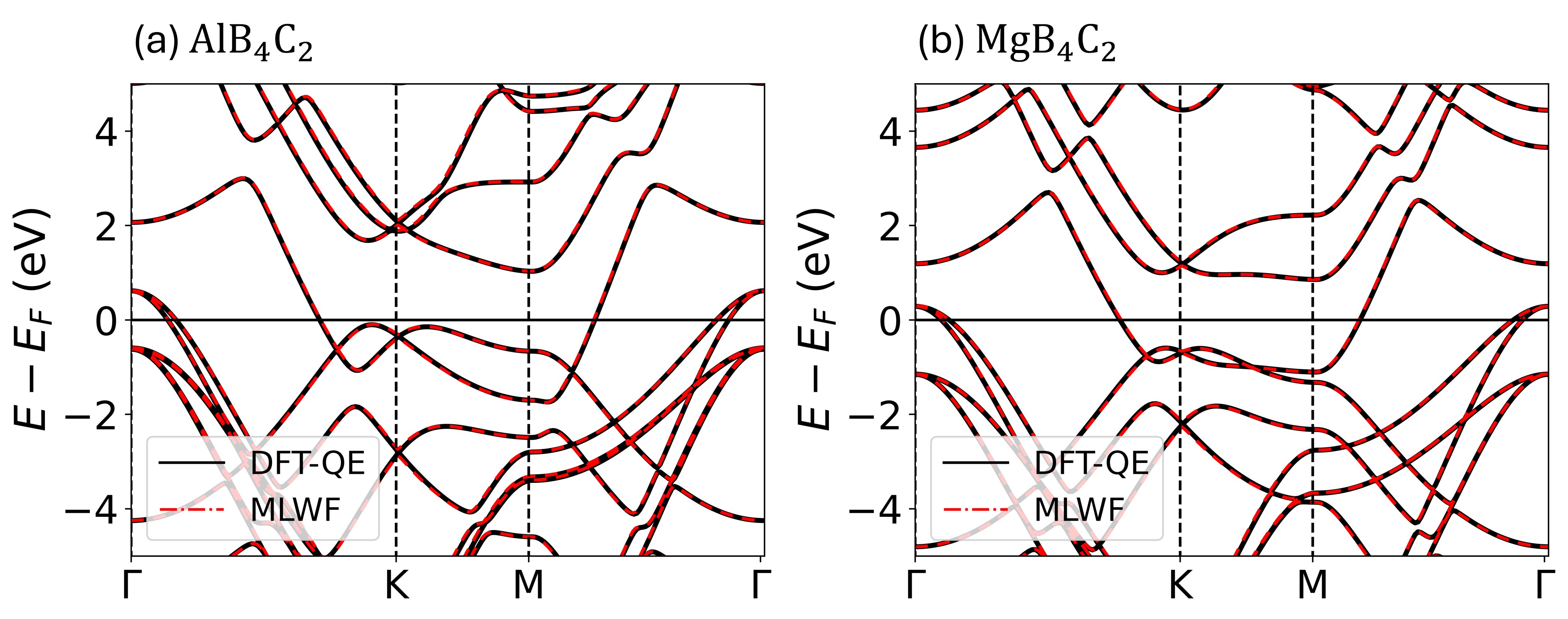} 
    \caption{Comparison of the electronic band structures obtained from direct DFT calculations in \textsc{Quantum ESPRESSO} (black solid lines) and those interpolated using Maximally Localized Wannier Functions (MLWF, red dash-dotted lines) for monolayer (a) $\text{AlB}_4\text{C}_2$, (b) $\text{CaB}_4\text{C}_2$, and (c) $\text{MgB}_4\text{C}_2$. The horizontal grey line represents the Fermi level ($E_F$), which is set to zero.}
    \label{fig:Fig-Appendix-A1}
\end{figure}

Figure~\ref{fig:Fig-Appendix-A1} compares the electronic band structures calculated explicitly using density functional theory (DFT) with those obtained via MLWF interpolation. The nearly perfect overlap between the exact DFT bands (black solid lines) and the interpolated bands (red dash-dotted lines) across a wide energy window confirms the reliability and precision of the Wannierization procedure for all three $M\text{B}_4\text{C}_2$ systems.

\bibliography{references}

\end{document}